\documentclass[aps,rmp,twocolumn,groupedaddress,floatfix]{revtex4-1}
\pdfoutput=1
\usepackage{graphicx}
\usepackage{amsmath}
\usepackage{amssymb}

\newcommand{\fig}[1]{Fig.~\ref{#1}}

\newcommand{\sseed}{s_0}
\newcommand{\smean}{\bar s}
\newcommand{\fpeak}{f_{\textrm{peak}}}
\newcommand{\sdown}{s_{\textrm{down}}}
\newcommand{\tildes}{\bar s}
\newcommand{\Ub}{U_b}
\def \so{s_0}
\def \fo{f_0}

\begin{document}
\renewcommand{\baselinestretch}{1.0}

\title{The Fates of Mutant Lineages and the Distribution of Fitness Effects of Beneficial Mutations in Laboratory Budding Yeast Populations}
\author{Evgeni M. Frenkel$^{1,2}$}
\author{Benjamin H. Good$^{2,3}$}
\author{Michael M. Desai$^{1,2,3,4}$}
\affiliation{
\mbox{${}^1$ Program in Biophysics, Harvard University, Cambridge, MA, USA}\\ 
\mbox{${}^{2}$ FAS Center for Systems Biology, Harvard University, Cambridge, MA, USA} \\ 
\mbox{${}^3$ Department of Physics, Harvard University, Cambridge, MA, USA} \\ 
\mbox{${}^4$ Department of Organismic and Evolutionary Biology, Harvard University,}\\ 
\mbox{Cambridge, MA, USA} } 

\begin{abstract}
The outcomes of evolution are determined by which mutations occur and fix. In rapidly adapting microbial populations, this process is particularly hard to predict because lineages with different beneficial mutations often spread simultaneously and interfere with one another's fixation.  Hence to predict the fate of any individual variant, we must know the rate at which new mutations create competing lineages of higher fitness. Here, we directly measured the effect of this interference on the fates of specific adaptive variants in laboratory {\it Saccharomyces cerevisiae  } populations and used these measurements to infer the distribution of fitness effects of new beneficial mutations. To do so, we seeded marked lineages with different fitness advantages into replicate populations and tracked their subsequent frequencies for hundreds of generations. Our results illustrate the transition between strongly advantageous lineages which decisively sweep to fixation and more moderately advantageous lineages that are often outcompeted by new mutations arising during the course of the experiment. We developed an approximate likelihood framework to compare our data to simulations and found that the effects of these competing beneficial mutations were best approximated by an exponential distribution, rather than one with a single effect size. We then used this inferred distribution of fitness effects to predict the rate of adaptation in a set of independent control populations. Finally, we discuss how our experimental design can serve as a screen for rare, large-effect beneficial mutations.
\end{abstract}
\date{\today}
\maketitle

\section{Introduction}

Evolutionary adaptation is driven by the accumulation of beneficial mutations. There are two basic questions one can ask about this process. First, what are the set of mutations available to the population? That is, what is the overall mutation rate, $U$, and the distribution of fitness effects, $\rho(s)$, of new mutations? Second, what is the fate of those mutations that occur? In other words, how does the frequency of each mutation change over time until it eventually fixes or goes extinct?

When beneficial mutations are rare, these two questions are independent. Mutations of a given fitness effect, $s$, occur at rate $U \rho(s)$. The fate of each mutant is then decided entirely on its own merits: it increases in frequency (or is lost due to random drift) at a rate commensurate with its selective effect. Experiments, however, have shown that even for modestly sized laboratory populations of viruses and microbes, multiple beneficial mutations often spread simultaneously and interfere with one another, an effect known as {\it clonal interference} (\citet{de1999diminishing, miralles1999clonal, joseph2004spontaneous, perfeito2007adaptive, desai2007speed, kao2008molecular, lee2013synchronous}; see \citet{sniegowski2010review} for a recent review). This means that the fate of each beneficial mutation depends not only on its own effect, but also on its interactions with the rest of the variation in the population \citep{lang2011genetic, lang2013pervasive}. In this regime, the mutation rate and the distribution of fitness effects of beneficial mutations (the DFE, $\rho(s)$) controls the availability of competing mutations, which then play an important role in determining the fate of each new beneficial mutation \citep{gerrish1998fate, good2012distribution}.

These factors highlight the importance of the DFE as a central parameter in adaptation, determining which new mutations occur and influencing their subsequent fate. Some theoretical work has argued that the DFE will typically be exponential \citep{gillespie1983simple, orr2003distribution}. However, this is fundamentally an empirical question, and in principle the details of the DFE could be highly system-specific. There has thus been extensive experimental effort devoted to measuring the DFE of beneficial mutations in a variety of laboratory populations (\citet{imhof2001fitness,rozen2002fitness, sanjuan2004distribution, barrett2006mutations, kassen2006distribution, burch2007experimental, perfeito2007adaptive, rokyta2008beneficial,maclean2009distribution,  bataillon2011cost,mcdonald2011distribution}; a separate literature has used population genetic methods to infer the DFE in natural populations, reviewed by \citet{keightley2010can}).

Experimental efforts to measure the DFE of beneficial mutations in laboratory populations have largely taken one of two complementary approaches. The first approach is to isolate mutants and directly assay their fitness. The difficulty with this method is that beneficial mutations are rare, so many clones must be screened to isolate comparatively few beneficial mutations \citep{sanjuan2004distribution, kassen2006distribution}. To avoid this difficulty, some studies have imposed a harsh selection and studied the survivors, which by definition must have a beneficial mutation \citep{maclean2009distribution, mcdonald2011distribution}. However, this approach is limited to harsh and typically narrow stresses (e.g. treatment with antibiotic), which may not be representative of adaptation to other conditions.

The second common experimental approach is to track the frequencies of genetic markers over time, and use the resulting dynamics to infer the underlying DFE. Such ``marker divergence'' experiments typically use two or more strains that differ by a single neutral genetic marker which can be easily tracked through time (e.g. antibiotic resistance or a fluorescent reporter). These strains are mixed, usually in equal proportions, and allowed to evolve in competition. The changes in frequencies of the neutral markers then reflect subsequent beneficial mutations that occur in one or the other genetic background \citep{novick1950experiments, atwood1951periodic, helling1981maintenance, paquin1983frequency, adams1986structure,  imhof2001fitness, barrett2006mutations, de2006clonal, perfeito2007adaptive, hegreness2006equivalence, kao2008molecular, barrick2010escherichia, lang2011genetic}. Inferring the DFE from such data typically requires estimating the fitness effects of many mutations from the dynamics of relatively few markers, which is naturally quite difficult \citep{hegreness2006equivalence, pinkel2007analytical, zhang2012estimation, illingworth2012method, de2013abc}. In principle, this difficulty could be removed by reducing the population size to such a degree that only one or zero beneficial mutations usually arise in each population \citep{perfeito2007adaptive}. However, this requires careful tuning of the population size, in order to make it small enough to minimize multiple mutations but also large enough to ensure that many replicates acquire a beneficial mutation.

Here, we introduce a twist on the traditional design of marker divergence experiments that produce dynamics more directly revealing of the underlying DFE. Rather than using neutral markers, we tracked the frequencies of marked lineages with a fitness \emph{advantage} relative to a reference strain. We seeded these marked lineages at low frequency into populations of the reference, so that their subsequent dynamics are reflective of the fates of beneficial mutations with a particular selective advantage. Since the DFE controls the availability of competing mutations and hence the likelihood of clonal interference, we can exploit the observed fates of seeded lineages to infer the DFE. Using lineages with different fitness advantages enabled us to probe different corresponding portions of the DFE. This approach is particularly suited to infer those aspects of the DFE that are most important in determining the fates of new beneficial ``driver'' mutations, e.g. the high-fitness tail, which is otherwise hard to measure directly. In the process, we also directly measured how clonal interference alters a key quantity in adaptation: the fixation probability of a beneficial mutation as a function of its fitness effect.

\section{Materials \& Methods}
\subsection{Strains}\label{sec:Strains}

All strains used in this study were derived from the base strain DBY15084, a haploid \emph{S. cerevisiae} strain derived from the W303 background with genotype MAT{\bf a}, ade2-1, CAN1, his3-11 leu2-3, 112, trp1-1, URA3, bar1$\Delta$::ADE2, hml$\alpha\Delta$::LEU2. Each experimental population included a resident and a seeded lineage. The resident lineage was DBY15108, a derivative of DBY15084 in which the fluorescent protein ymCherry was integrated at the URA3 locus \citep{lang2011genetic}. The seeded lineages were descendants of strain DBY15104 isolated from timepoints of an earlier long-term evolution experiment \citep{lang2011genetic}. To allow us to track their frequency using flow cytometry, we amplified a pACT1-ymCitrine pTEF-HISMX6 cassette from plasmid pJHK043 (provided by John H. Koschwanez) and integrated it at the HIS locus using oligos oGW137 (5'-TTGGTGAGCGCTAGGAGTC-3') and oGW138 (5'-TATGAAATGCTTTTCTTGTTGTTCTTACG-3') provided by Gregg Wildenberg. From this pool of transformants, we selected strains EFY11-17 based on fitness assays described below.

\subsection{Experimental procedures}
\label{sec:ExpProceds}

\begin{figure*}
\includegraphics[width=7in]{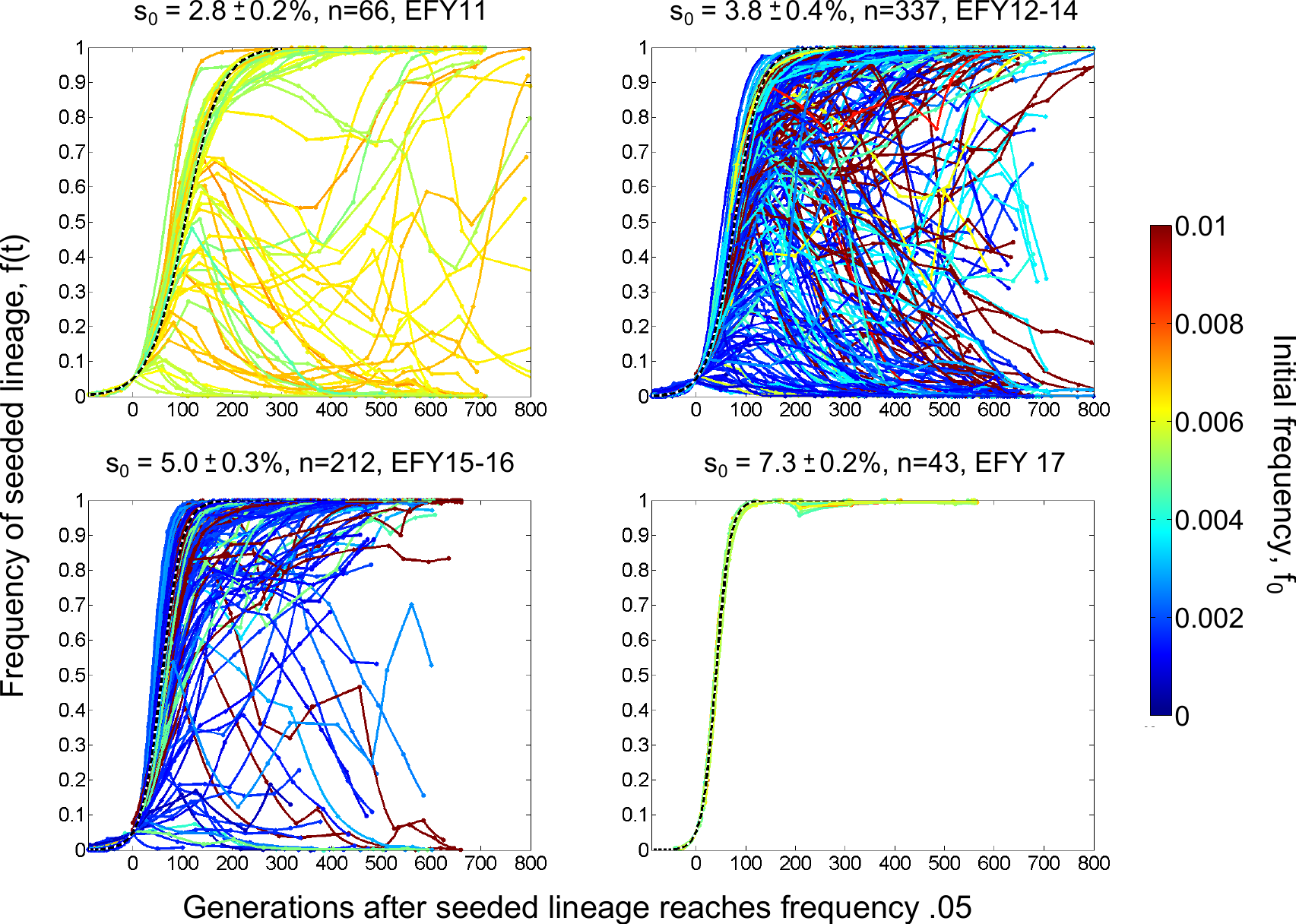}
\caption{{\bf Trajectories of seeded lineages.} Each line represents the frequency over time of a marked lineage with fitness advantage $\sseed$ seeded into a replicate resident population.  Colors correspond to the initial frequency $f_0$ of the seeded lineage according to the legend at right. Time is measured in generations, with $t=0$ defined as the time at which each trajectory reached frequency $0.05$. The dashed curves show the expected trajectories in the absence of new beneficial mutations (i.e. without clonal interference). (Note that the seeded lineages for $\sseed \approx 4\%$ and  $5\%$ consisted of multiple strains; see Methods). \label{fig:allrawdat}}
\end{figure*}

To obtain seeded lineage strains with a range of fitnesses, we isolated a large number of evolved clones and assayed their fitnesses as described in \citet{lang2011genetic}. Briefly, this protocol is to mix each strain in roughly equal proportion with a reference strain that bears a different fluorescent reporter, propagate these mixed populations for $30$ generations, and measure the ratio of the strains at generations $10$ and $30$ using flow cytometry. Relative fitness was calculated as $s=(1/20)\cdot \log($final ratio$/$initial ratio$)$. From among these clones, we chose EFY11-17 to use as seeded lineages and remeasured their fitnesses in 10 replicates. These additional assays showed that strains EFY12-14 and EFY15-16 had indistinguishable fitnesses, and so for the purposes of analysis, strains EFY11, EFY12-14, EFY15-16 and EFY17 were respectively grouped into the fitness classes indicated in Fig.~\ref{fig:allrawdat}.

To begin the evolution experiment, we grew up an individual resident clone to saturation in 3mL of standard growth media (YPD supplemented with 100 $\mu$g/mL ampicillin and 25 $\mu$g/mL tetracyclin). We transferred 128$\mu$L of this culture into each well of a 96 well-plate, diluted these cultures  $2^{10}$-fold into twelve 96-well plates containing fresh media, allowed these cultures to grow for 10 generations, and froze them at -80$^o$C in 15\% glycerol. Later, these plates were thawed and propagated  for 30 generations (as described below) to re-acclimate them to this environment. In parallel, we prepared the seeded clones in the same fashion. We then mixed seeded and resident populations to found a total of 1044 populations in twelve 96-well plates (see Table S1). These populations were propagated at $30^o$C in 128$\mu$l YPD per well and diluted every 24 hours by a factor of $2^{10}$ into new plates containing fresh media. This corresponds to an effective population size $N_e \approx 10^5$ \citep{lang2011genetic, wahl2002evaluating}. Each plate contained a set of 9 empty wells as cross-contamination controls. All control wells remained sterile throughout the experiment except for two accidents involving plate mixing.  This contamination was resolved by restarting from glycerol stocks of an earlier time point. Transfers were carried out using a Biomek FX pipetting robot.

At approximately 50-generation intervals, seeded lineage frequencies were measured using flow cytometry. In particular, BD Biosciences Fortessa and LSR-II flow cytometers with high-throughput plate samplers counted $\sim$100,000 cells per population for the initial time point and $\sim$30,000 cells per population for time points thereafter. Repeated measurement of populations and blanks indicated that roughly $\sim$100 cell counts per sample were carry-over from previous samples. Therefore the uncertainty in frequency at the first time point was $\sim$0.1\% and $\sim$0.5\% thereafter. These raw data were processed in FlowJo version 9.2. All processed data are provided in Table S1.

We also assayed the fitness of 16 additional control populations founded with only the resident strain. To do so, these populations were thawed from frozen-archive plates, each was duplicated into 4 replicates, these were propagated for 30 generations to acclimate them, and then their fitness was assayed as described above.

Note that 386 of the populations were later excluded from analysis, leaving a total of 658 replicate populations, apportioned among the seven seeded lineages and controls as described in Table S1. In 232 of these, frequency dependent selection emerged. We identified these by first investigating 15 populations in which lineages co-existed at constant proportion for hundreds of generations. We found that this co-existence was maintained by frequency dependent selection exclusively in populations having a characteristic pellet morphology, so we excluded from analysis all populations that also had this morphology. In the other cases, the initial frequency of the seeded lineage was so low that it could not be precisely determined or extinction due to drift was common. To exclude these without biasing the statistics of trajectories, we chose a cut-off for the initial frequency of each seeded strain such that in all replicates in which the initial frequency was above the cut-off, the seeded lineage rose to at least $5\%$. All replicates below the cut-off were excluded.

\section{Results}

\subsection{Tracking the fates of seeded lineages}

Any beneficial mutation creates a new lineage that is more fit than the genetic background in which it arose. To systematically study the fates of such lineages, we prepared a set of fluorescently labeled haploid budding yeast strains (the \emph{seeded lineages}) with measured fitness advantages, $\sseed$, of approximately 3, 4, 5 and 7\% relative to a closely related but separately labeled reference strain. We founded 658 replicate populations of the reference (the \emph{resident}), and introduced one of the seeded lineages at low frequency into each replicate population. We propagated these populations asexually in batch culture for hundreds of generations at an effective population size of $N_e \approx 10^5$, measuring the frequency of the seeded lineage in each population approximately every 50 generations (see Methods). This allowed us to track the fate of the seeded lineages over time, as illustrated in \fig{fig:allrawdat}.

Each seeded lineage was introduced at an initial frequency $f_0$ large enough that genetic drift is expected to be weak relative to natural selection (i.e. $f_0 \gg \frac{1}{Ns}$). In the absence of additional mutations, this implies that the frequency $f(t)$ of each seeded lineage should increase deterministically according to the logistic equation, $f(t) = \frac{f_0 e^{st}}{1+f_0 (e^{st}-1)}$. This expectation is indicated by the dashed curves in \fig{fig:allrawdat}. As is apparent from the figure, most seeded lineages initially conformed to this expectation (the exceptions are lineages whose initial frequencies were only several-fold greater than $\frac{1}{Ns}$, which is low enough that genetic drift could partially reduce their initial rate of increase). Subsequently, many lineages diverged into a variety of qualitatively distinct fates. Since both genetic drift and measurement errors are expected to be small relative to this divergence (see Methods), the variation in the fates of seeded lineages indicates that their relative fitnesses were modified by new beneficial mutations arising during the experiment.

\subsection{Fates of seeded lineages reflect supply of competing beneficial mutations}

\begin{figure*}
\begin{center}
\includegraphics[width=6in]{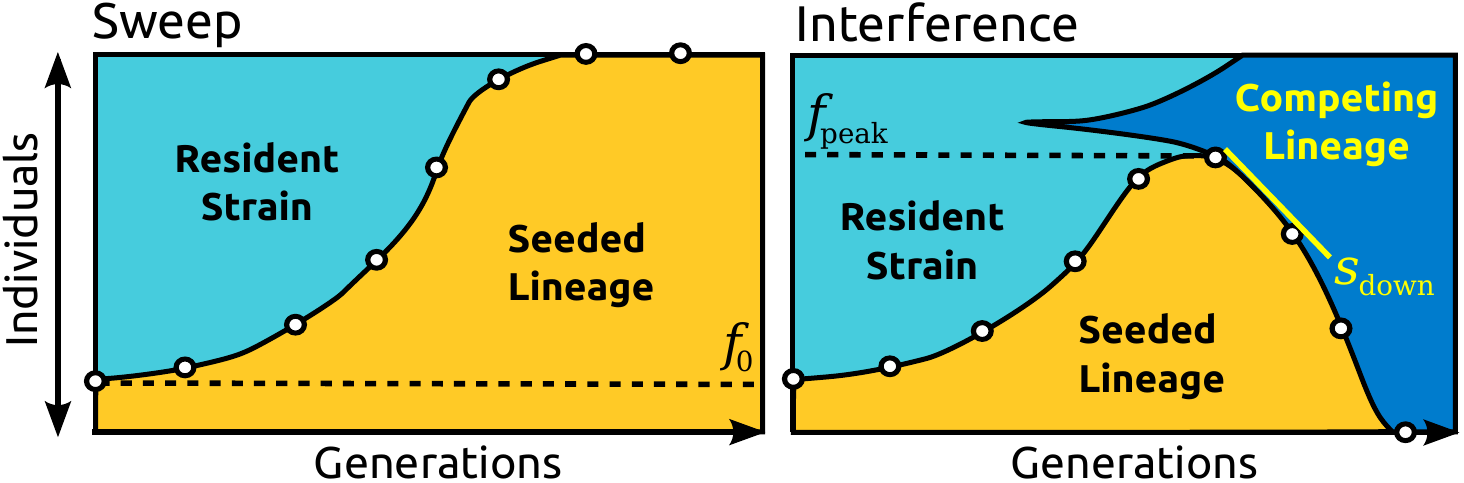}
\caption{{\bf The fates of seeded lineages. } We classified the trajectory of each seeded lineage according to whether it increased monotonically to fixation (a selective sweep, shown left) or peaked and subsequently declined in frequency (clonal interference, shown right). Each clonal interference event implies that the resident adapted fast enough to overtake the seeded lineage in fitness. These cases were further classified by the seeded lineage's peak frequency, $\fpeak$, and relative fitness after this peak, $\sdown$, as indicated in the schematic at right.  \label{fig:muller} \label{fig:trajInterp}  }
\centering
\end{center}
\end{figure*}

\begin{figure}
\begin{center}
\includegraphics[width=3.4in]{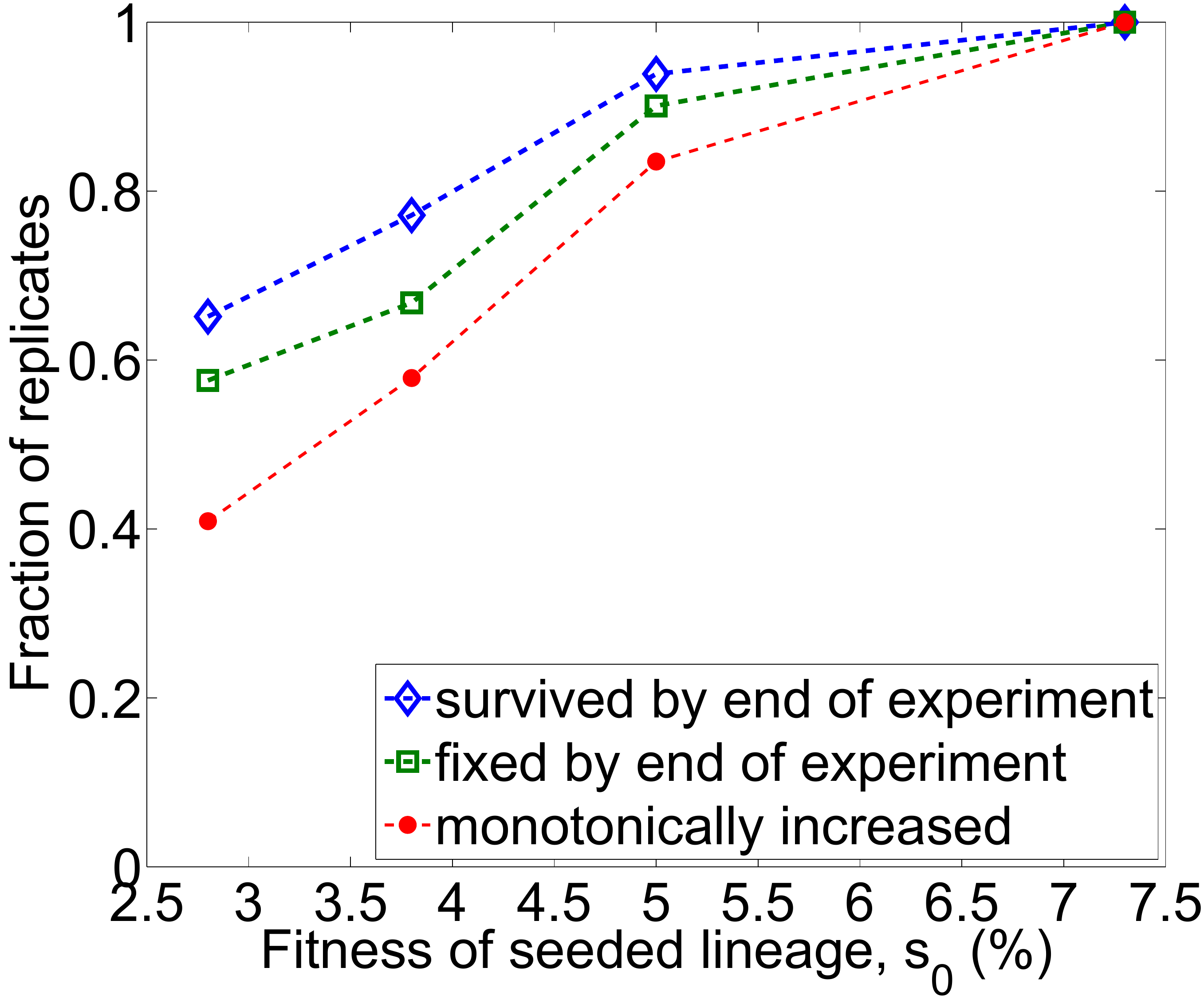}
\caption{{\bf The fates of seeded lineages as a function of their fitness.} We show the fraction of replicate populations in which the seeded lineage had the indicated fate.\label{fig:pFix} }
\centering
\end{center}
\end{figure}

The trajectory of each seeded lineage provides information about the beneficial mutations that did (or did not) arise within the competing resident population. Consider for example the case where a seeded lineage of fitness $\sseed$ peaks and then declines in frequency. This reflects a clonal interference event, where one or more new beneficial mutations in the resident population create a competing lineage with fitness greater than $\sseed$ (see \fig{fig:muller}). By considering the range of outcomes in replicate populations, we can estimate the probability of these events (\fig{fig:pFix}). A higher probability of clonal interference implies a larger supply of beneficial mutations that can generate successful competing lineages.

Comparing the fates of seeded lineages of different fitnesses provides additional insight into the  mutations responsible for clonal interference. For example, the seeded lineage with fitness advantage $\sseed = 7\%$ always swept to fixation without any detectable deviation from the expectation in the absence of interference. In contrast, the lineage with $\sseed = 5\%$ swept in 84\% of replicates. Together, these two results suggest that clonal interference in the $\sseed = 5\%$ case was primarily due to beneficial mutations in the resident that created competing lineages with fitness advantages between 5 and 7 percent. Extending this logic, comparing the fates of seeded lineages with $\sseed = 5$, 4, and 3 percent provides information about the probabilities that beneficial mutations create competing lineages of fitness between 4 and 5 percent and between 3 and 4 percent.

While this intuition is straightforward, quantitative inference of the DFE requires us to connect the rates of individual mutations with the fitnesses of competing lineages. This is complicated because competing lineages may often contain multiple beneficial ``driver'' mutations. In addition, beneficial mutations may also arise in seeded lineages, despite their initially much smaller population sizes. To fully account for these effects, we now introduce a computational method for inferring the DFE.

\subsection{DFE inferred from seeded lineage dynamics}

We implemented an approximate likelihood method which uses information from the shapes of the trajectories of seeded lineages to infer the DFE of beneficial mutations. Any particular trajectory only carries information about the beneficial mutations that rose to significant frequency in that population (i.e. the ``contending'' mutations; \citet{rozen2002fitness}), but by modeling the trajectories of many populations together, we can learn about the overall distribution of possible beneficial mutations for the strains in our experiment. In order to make this inference tractable, we limited ourselves to single-parameter DFE shapes characterized by an average fitness effect $\tildes$ and beneficial mutation rate $U_b$. For concreteness, we considered three canonical distributions commonly used in the literature: an exponential DFE, $\rho_{\textrm{exp}}(s) = \frac{1}{\tildes} e^{-s/\tildes}$, a uniform DFE $\rho_{\textrm{unif}}(s) =\textrm{Heaviside}(2\tildes-s)/(2\tildes)$, and a $\delta$-function DFE  where all beneficial mutations have the same fitness effect, $\rho_{\delta}(s) = \delta({s - \tildes})$. We explain the significance of these choices in the Discussion.

To compute the likelihood of particular DFE parameters, we ran forward-time simulations of the experiment and estimated the likelihood as the fraction of replicate simulations that matched the data (see Appendix). In principle, we could use the complete trajectory of each seeded lineage for this comparison, identifying a match between simulations and data whenever the two were identical. However, in practice this was not computationally tractable. Instead, we focused on two features of the dynamics: the first peak frequency, $\fpeak$, of each seeded lineage (binned into quartiles, including fixed lineages) and the rate at which the seeded lineage declined in frequency following this peak, $\sdown$ (binned into 2\% intervals). These are illustrated in \fig{fig:muller}. We chose to focus on these two quantities because we expect them to be particularly sensitive to the DFE: $\fpeak$ indicates how quickly a competing lineage arose in the resident population, while $\sdown$ measures how much the relative fitness of the resident population increased in this time. In addition, this focus on early-time dynamics ensures that most relevant mutations occur in the resident (due to its initially much larger population size), minimizing the effects of potential differences in the DFEs of the seeded genotypes.

\begin{figure*}
\begin{center}
\includegraphics[height=4.3in]{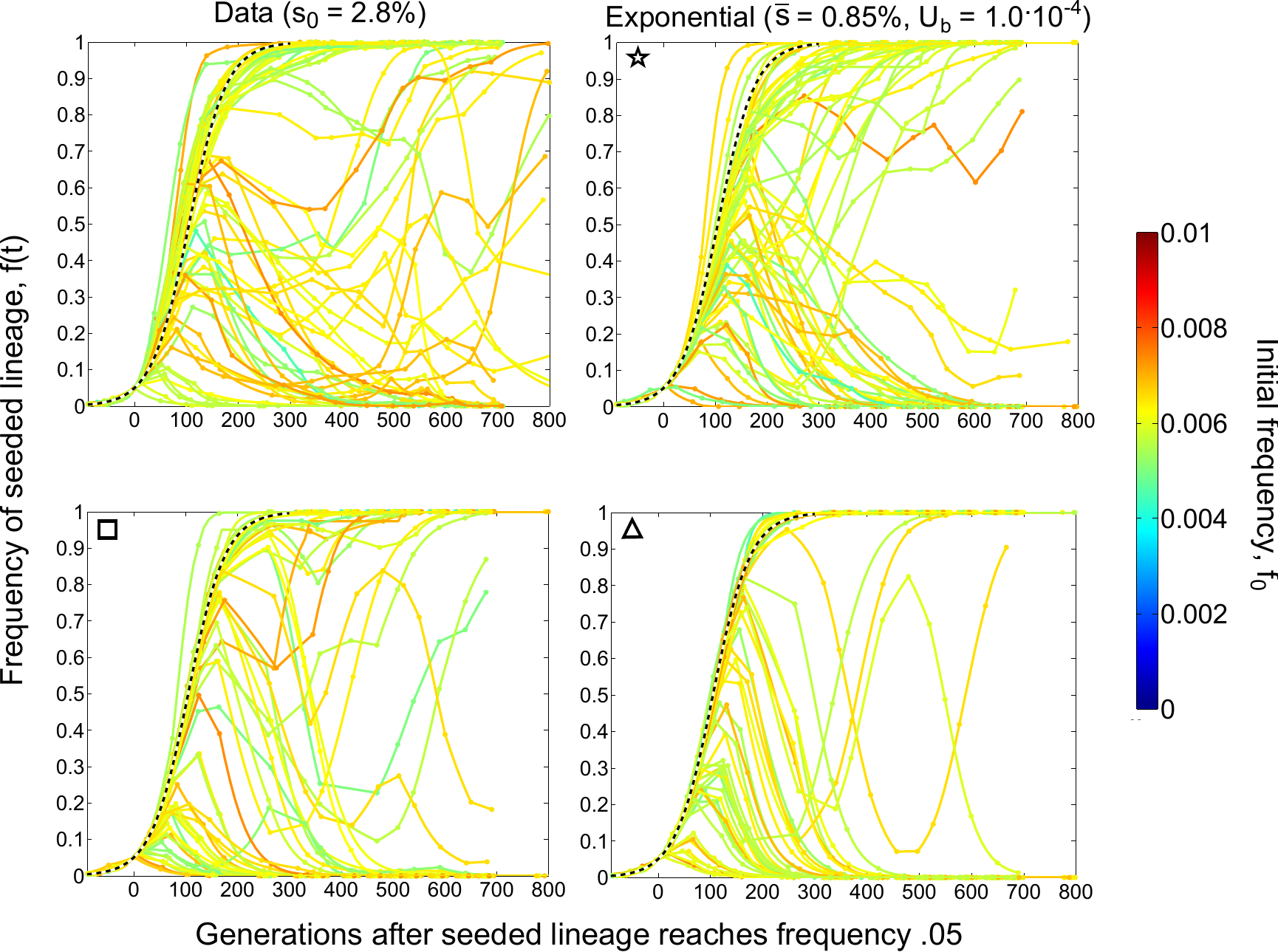}
\caption{{\bf Lineage dynamics data and simulations for $\mathbf{\so = 2.8\%}$.} Each panel shows the trajectories of seeded lineages with initial fitness $\so=2.8\%$ as observed in the experiment (top left) and as reproduced by simulations assuming the DFE parameters indicated above each panel. \label{fig:linDynamicsDatAndSims1}}
\centering
\end{center}
\end{figure*}

For the three considered DFE shapes, we identified the most-likely parameters $U_b$ and $\tildes$ by scanning a grid of candidate values. These parameters are shown in \fig{fig:DFEsInferred}, along with confidence bounds estimated by bootstrapping (see Appendix). For each of these most-likely parameters, we show simulations of the $\sseed=3\%$ seeded lineage trajectories in \fig{fig:linDynamicsDatAndSims1} and for the $\sseed = 4, 5 \mbox{ and } 7\%$ lineages in Figs.~\ref{fig:linDynamicsDatAndSims2}-3. Using a likelihood ratio test, we found that the exponential DFE provided a significantly better fit to the data than either the $\delta$-function ($p<10^{-4}$) or uniform distribution ($p<10^{-4}$) and that the uniform provided a better fit than the $\delta$ ($p<10^{-4}$). 

Since the seeded lineage with $\sseed = 7.3 \%$ always swept to fixation, indicating that larger-effect mutations must be rare, we checked whether truncating the high fitness end of the exponential DFE would improve its fit to the data. To do so, we considered an exponential DFE truncated at $7.3 \%$ and performed the same inference and statistical tests as above. We found that this truncated exponential provided a better fit to the data, but not significantly so ($p>0.08$, likelihood ratio test). We also checked whether truncating the low-fitness end of the exponential would affect its fit to the data. We varied this truncation and found that, for the inferred exponential DFE parameters, discounting mutations with fitness effects below $2.1\%$ improved these parameters' fit to the data, but only marginally so. This indicates that the seeded lineages were not strongly affected by mutations with fitness effects below $\sim2\%$.

\begin{figure*}
\begin{center}
\includegraphics[width=7in]{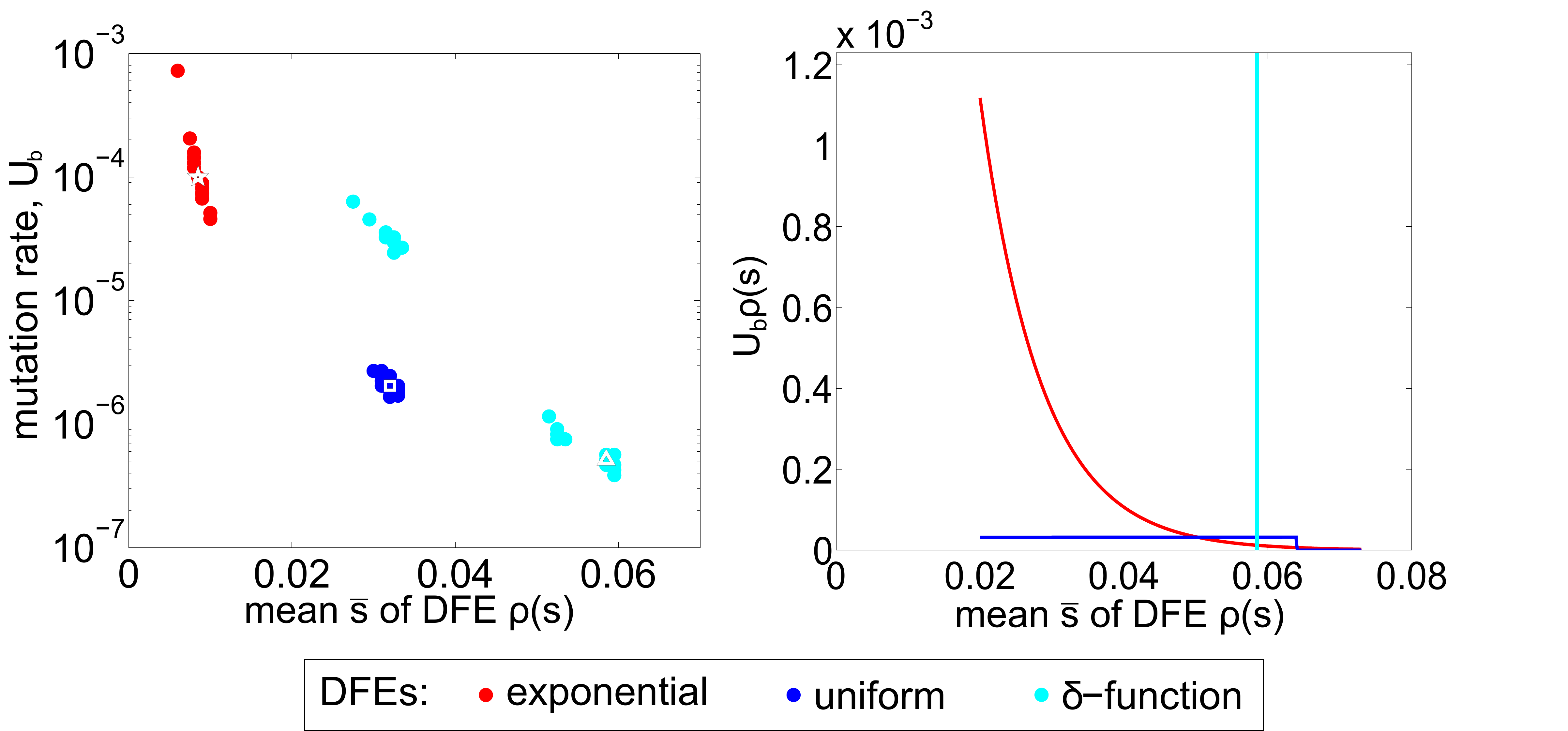}
\caption{{\bf Inferred DFE parameters.} Left: The most-likely parameters $U_b$ and $\tildes$ for DFEs $\rho(s)$ given by exponential (star), uniform (square) and $\delta$-function (triangle) distributions. Shaded circles indicate 1\% confidence ranges of these parameters as estimated by bootstrapping (see Appendix). Right: The shapes of these distributions shown for $s\geq2\%$ given their most-likely parameters. \label{fig:DFEsInferred}
}
\centering
\end{center}
\end{figure*}

\subsection{Measurements of adaptation rate corroborate DFE inference}

\begin{figure}
\begin{center}
\includegraphics[width=3.4in]{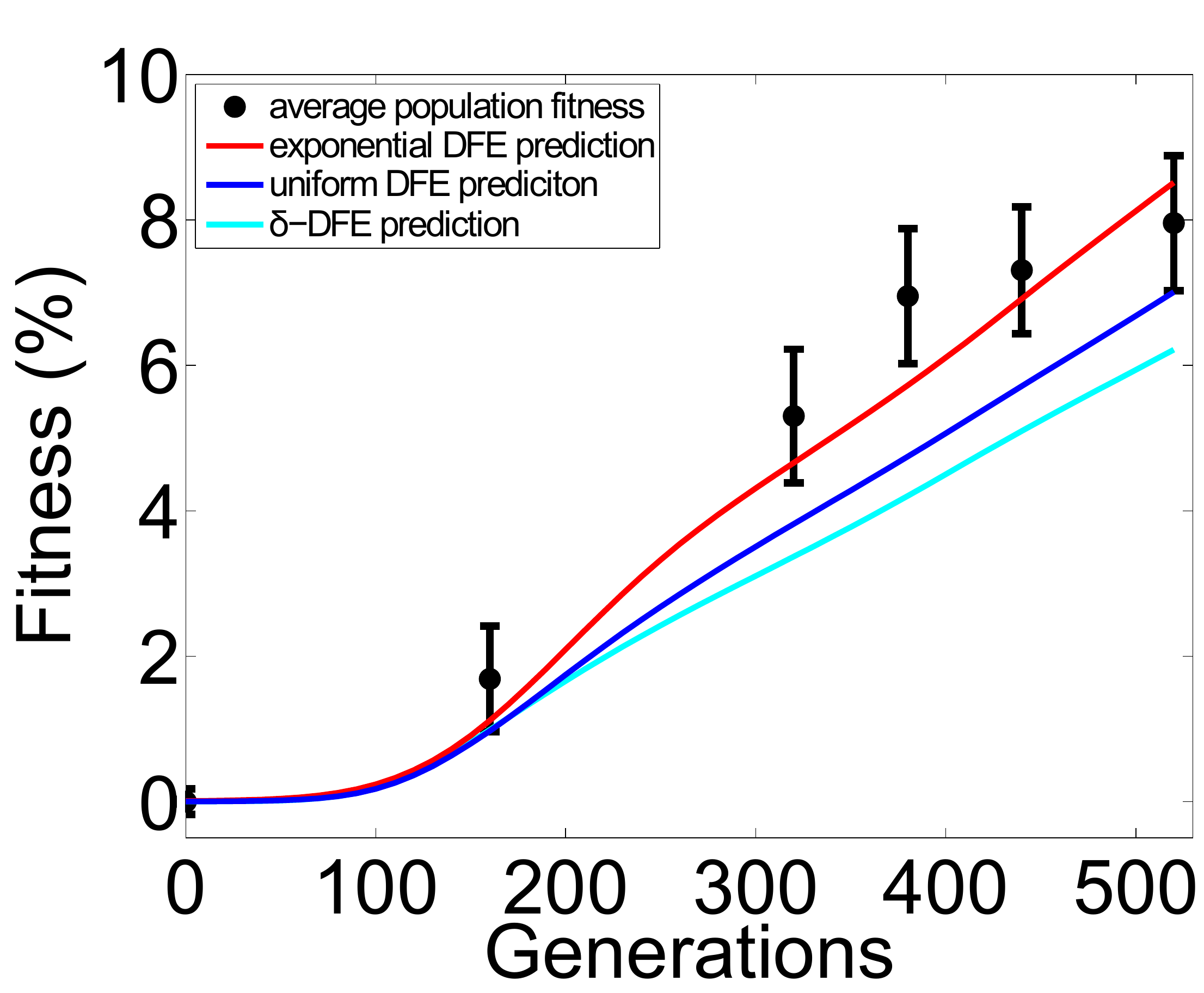}
\caption{{\bf The rate of adaptation.} The average fitness over time of 16 experimental control populations ($\pm 1$~s.e.m.) is shown in black. Solid curves are the predictions of these data given the most-likely exponential, uniform, and $\delta$-function DFEs. The fitnesses of individual populations are shown in \fig{fig:adapRateDatVsimSI}. \label{fig:adapRateDatVsim}}
\centering
\end{center}
\end{figure}

In addition to determining the dynamics of seeded lineages, the DFE determines the rate of adaptation. Thus to test our inferences, we measured the changes in fitness over time of 16 control populations that consisted of the resident strain alone. We compared the average fitness of the control populations with the predictions of the most-likely exponential, uniform and $\delta$-function DFEs. As seen in \fig{fig:adapRateDatVsim}, the inferred exponential is fairly accurate in predicting these data, whereas the uniform and $\delta$-function are less so. 

Throughout our analysis, we have implicitly assumed that the DFE remained the same across all genotypes in the experiment, which implies that the fitnesses of populations should increase linearly on average after some initial transient. In contrast, the rate of adaptation slowed after generation 380 ($p < 3\cdot10^{-3}$, see Appendix), which is reminiscent of declines in adaptation rate commonly observed in other evolution experiments \citep{elena2003evolution}. Fortunately, we based our DFE inference on the early features of seeded lineage dynamics, most of which transpired prior to this time. Thus the change in adaptation rate is not inconsistent with our method.

\section{Discussion}

Interest in the DFE stems from a desire to know what beneficial mutations are available and which of these drive adaptation. In asexual populations, the DFE also determines the distribution of competing mutations and the frequency of clonal interference. Here, we have described a simple experiment which exploits this connection in order to infer the DFE in experimental populations of \emph{S. cerevisiae}. By introducing lineages with different fitnesses and tracking their subsequent dynamics, we inferred the DFE from the statistics of observed interference events. In the process, we directly observed how initial fitness advantages and clonal interference jointly influence the fixation or loss of adaptive lineages.

Previous experimental work has analyzed several other cases where an introduced lineage is outcompeted by a less-fit resident population \citep{gifford2013evolutionary, waite2012adaptation}. Unlike our experiment, these earlier studies focus on the fates of a few key mutations (e.g., antibiotic resistance or microbial ``cheaters'') without attempting to infer the underlying DFE. Nevertheless, our results complement this earlier work by showing the transition between fitness effects that are susceptible to clonal interference and those that decisively sweep to fixation, which has previously been studied theoretically \citep{schiffels2011emergent, neher2011genetic, good2012distribution}. In our system, this transition occurs when the fitness of the seeded lineage is about $5$ percent, which represents a critical effect size required for a mutation to drive adaptation. Of course, in natural populations some adaptive variants may arise in populations with substantial \emph{standing} fitness variation, rather than the homogeneous resident populations employed here. In this case, the transition between mutations that sweep and those that experience interference is determined both by the DFE and by the distribution of fitnesses in the resident population. Further work is needed to address this situation.

Our computational inference method allowed us to distinguish between three representative DFE shapes: exponential, uniform, and  $\delta$-function (in which all mutations have the same effect). These represent idealized approximations to the actual DFE, and it is likely that a larger number of replicates or more sophisticated computational techniques could produce other DFE shapes with a significantly better fit. Yet one cannot continue this process indefinitely without reaching a point where further determination of the fine-scale DFE becomes irrelevant for any particular application. In the end, certain features of the DFE matter for predicting certain aspects of the evolutionary process, and the required level of resolution is ultimately determined by the aspect of adaptation one wishes to study. Our present experiment, which focuses on the fates of advantageous mutants, provides a concrete illustration of this principle. Previous work has suggested that the dynamics of adaptation can be summarized by a single characteristic fitness effect, with a magnitude that depends on the actual DFE and the level of clonal interference within the population \citep{hegreness2006equivalence, desai2007beneficial, good2012distribution}. By rejecting the $\delta$-function and uniform DFEs in favor of the exponential, we have shown that this assumption breaks down when one considers more detailed features of the lineage trajectories. 

Given these caveats, the DFE that we inferred is worth pondering. We estimated an exponential distribution with mean $\tildes = 0.85\%$ and total beneficial mutation rate $U_b = 1.0\cdot10^{-4}$. Our modeling indicated that of these mutations, only those with effects greater than $2\%$ affected the fates of seeded lineages, and that these mutations are predicted to arise at a rate of order $10^{-5}$ per individual per generation. If one assumes a per-genome point mutation rate of roughly $4\cdot10^{-3}$ \citep{lynch2008genome}, this would imply that of order 1 in 1000 mutations confer a fitness advantage of two percent or more. This is consistent with past work in a related system \citep{desai2007speed}, and is also similar to DFEs reported for bacteria adapting to rich laboratory media \citep{perfeito2007adaptive, kassen2006distribution, wiser2013long}. In such permissive environments, other studies in yeast that have identified specific adaptive mutations report a mix of loss-of-function versus other kinds of beneficial mutations \citep{kao2008molecular, wenger2011hunger, jansen2005prolonged, lang2013pervasive,kvitek2013whole}. If a large fraction of beneficial mutations in our system are loss of function, and if roughly ten percent of spontaneous mutations in a gene cause loss of function \citep{lang2008estimating}, our results would suggest that about 1 in 100 genes are beneficial to disrupt. This is at least qualitatively consistent with direct measurements using the yeast deletion collection \citep{sliwa2005loss, bell2010experimental}. Together, these results illustrate how inferences from lineage dynamics can combine with other lines of evidence to help build a more complete picture of adaptation.

\begin{figure}
\begin{center}
\includegraphics[width=3.4in]{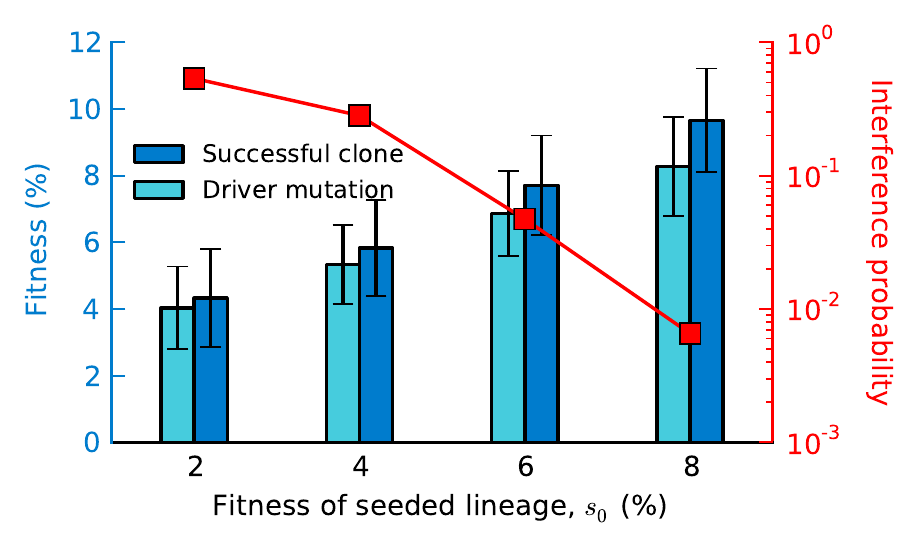}
\caption{{\bf A screen for beneficial mutations.} We simulated the evolutionary dynamics for a range of seeded lineages, and then simulated picking a single clone at random from the resident population immediately after a clonal interference event. The bars indicate the average fitness of this clone and its largest effect mutation ($\pm 1$ s.d., scale at left). We also show the fraction of replicate populations in our simulations in which clonal interference occurs (scale at right). The simulations assumed the most-likely exponential DFE inferred in the study. \label{fig:mutScreen}}
\centering
\end{center}
\end{figure}

Finally, we note that our experimental design has a potential practical application as a screen for beneficial mutations. Whenever a seeded lineage with fitness advantage $\sseed$ experiences clonal interference, the resident must contain a mutant lineage at appreciable frequency with fitness greater than $\sseed$. Thus, by picking clones from the resident immediately after a clonal interference event, we should in principle be able to isolate rare large-effect beneficial mutations. This is similar in spirit to earlier studies which used the dynamics of neutral markers to screen for adaptive clones (e.g. \citet{rozen2002fitness}). However, because our seeded lineages are more fit than the resident, we can screen for beneficial mutations with particularly large effects. Further, since the resident must quickly generate a competing lineage, our approach is more likely to find clones with  fewer mutations of larger effect rather than many of smaller effect, as well as limit the number of non-beneficial hitchhiking mutations. To illustrate this idea, we simulated seeded lineage trajectories and then simulated picking a clone from the resident population after observed clonal interference events. In \fig{fig:mutScreen}, we show the average fitness of each of these simulated clones and of the largest-effect mutation in each clone. As is apparent from the figure, it should be feasible to use this approach with a seeded lineage of the appropriate fitness to isolate large-effect beneficial mutations with specific fitness effects.

\section{Acknowledgments}
We thank Sergey Kryazhimskiy, Christopher S. Wylie, Andrew Murray, and Katya Kosheleva for useful discussions and comments on the manuscript; Melanie Muller, Gabriel Perron, John Koschwanez, and Gregg Wildenberg for help with strain construction; and Patricia Rogers for generous technical support of flow-cytometry. Simulations in this paper were performed on the Odyssey cluster of the Research Computing Group at Harvard University.

This work was supported by training grant GM831324 from the NIH and grant 1219334 from the NSF Physics of Living Systems graduate student network (E.M.F.), a National Science Foundation Graduate Research Fellowship (B.H.G.), and the James S. McDonnell Foundation, the Alfred P. Sloan Foundation, the Harvard Milton Fund, grant PHY 1313638 from the NSF, and grant GM104239 from the NIH (M.M.D.). 
 
\section{Appendix: Inferring the DFE}
\subsection{Simulations}\label{sec:simMeth}

For a given DFE, we simulated lineage trajectories using a forward-time algorithm designed to mimic the conditions of our experiment. Between each transfer, each cell expanded clonally for $10$ generations at a deterministic exponential growth rate $r=r_0+X$, where $X$ is the fitness of the cell relative to the resident ancestor strain. At the transfer step, the population was downsampled to $N_b = 10^4$ individuals with Poisson sampling noise. Mutations accumulate during the growth phase, but we assumed that they did not influence the fitness of the cells until the next transfer cycle. Thus, mutation was approximated by assuming that each individual has a probability $10 U_b$ of gaining a beneficial mutation at the end of a transfer step, with additive fitness effects drawn from the underlying DFE. In order to speed computation, we binned the fitnesses of individual cells into discrete fitness classes of width $\Delta s = 0.01\%$ for all simulations except those in \fig{fig:mutScreen}, which required information from individual mutations.

Each replicate simulation began at generation $t=0$ with a homogeneous seed population with initial fitness $\so$ and initial size $\fo N_b$, and a resident population of size $(1-\fo)N_b$, with $\so$ and $\fo$ as measured experimentally. The initial genetic composition of each resident population was obtained by simulating deterministic growth from a single-cell to $3 \times 10^8$ cells, followed by a Poisson dilution down to $N_b$ cells and four transfer cycles as described above. Simulated trajectories were then obtained by propagating the seeded lineage and the resident and recording the number of descendants of the seeded lineage at the same timepoints as the experiment, up to the time required for the fixation or first-peak used in the inference. Simulations for the rate of adaptation were carried out in a similar manner for populations consisting only of the resident (without the four transfer cycles prior to $t=0$). A copy of our implementation is available upon request.

\subsection{DFE parameter estimation} \label{sec:simStat}

To determine the likelihood of the data for a particular set of DFE parameters, the $650$ measured trajectories were partitioned into 13 classes such that the seeded lineages within each class shared the same initial fitness $\so$ and differed in their initial frequency $\fo$ at most 2-4 fold (Table S1). We classified each trajectory into one of $17$ bins of $(\fpeak,\sdown)$ values as described in the text. To estimate the relative probabilities of each of these bins, we simulated a large number of trajectories for each of the $13$ seeded lineage classes and recorded the fraction of times that each trajectory bin was observed. The total likelihood of the data for a given set of DFE parameters was then estimated as the product of the trajectory bin probabilities for each of the $650$ measured trajectories.

We determined the most-likely parameters for a particular DFE shape by scanning across a grid of $\Ub$ and $\smean$ values, which was locally resampled at finer resolutions until the most-likely parameters could be identified with a reasonable level of confidence. We first simulated a coarse grid of parameter values with a mean rate of adaptation between 0 and $5\%$ per 100 generations. We confirmed by visual inspection that the likelihood surface smoothly sloped toward the most-likely point identified in this coarse grid. We drew a rectangle around this peak and resampled points and adjusted the boundaries of this region until they satisfied the following criteria: (1) Any infinitesimal area of the region contained at least one point whose likelihood uncertainty (due to the finite number of simulated trajectories) was less than 0.5 log-likelihood units (LLU). Here, infinitesimal areas correspond to $10\%$ increments of $U_b$ and $0.1\%$ increments of $\smean$. We estimated the uncertainty in the likelihood using the Wilson confidence interval \citep{brown2001interval} and employed a minimum of $10^4$ simulated trajectories per parameter value. (2) Each infinitesimal area on the border of the peak region contained a point with likelihood at least 10 LLU below the peak and whose uncertainty was less than 0.5 LLU. Once these criteria were met, the most-likely parameters for the candidate DFE shape were estimated to be the grid point with the highest likelihood value. We estimated the confidence regions in \fig{fig:DFEsInferred} by re-fitting the most-likely parameters for $10^4$ bootstrapped datasets, which we obtained by resampling the observed trajectories with replacement in such a way that the total number of trajectories in each of the 13 trajectory classes was preserved. Fig.~\ref{fig:DFEsInferred} shows the scatter of parameters that were found to be most-likely for at least $1\%$ of these bootstrapped data sets.

\subsection{Statistical tests} \label{sec:statTests}

We used a standard likelihood ratio test to evaluate whether the most-likely exponential DFE provided a significantly better fit than the most-likely $\delta$-function DFE. To obtain the null distribution of the likelihood ratio, we simulated $10^4$ data sets using the most-likely $\delta$-function DFE and determined the most-likely parameters for each of these simulated datasets under the exponential and $\delta$-function DFEs as described above. We then estimated the $p$ value as the fraction of simulated data sets whose likelihood ratio was more extreme than the value obtained from the measured trajectories. A similar procedure was used to compare the exponential and truncated exponential DFEs, with the exponential DFE now taking the role of the null hypothesis.

To obtain an absolute measure of goodness-of-fit for the exponential and $\delta$-function DFEs, we used the estimated maximum likelihood as a test statistic and generated $10^4$ simulated datasets given the most-likely values of $U_b$ and $\smean$. We then estimated the $p$-value as the fraction of simulated datasets whose estimated maximum likelihood was lower than that of the actual data.

The significance of the slowdown in adaptation rate was assessed with a non-parametric bootstrap procedure. We generated $10^4$ bootstrapped datasets obtained by resampling the 16 populations with replacement, and for each of these, further resampling from the four fitness measurements at each timepoint. The null distribution for the change in adaptation rate, $\Delta v$, was obtained by calculating the change in adaptation rate in each bootstrapped dataset and subtracting the observed value from the original data. We then estimated the $p$-value as fraction of bootstrapped datasets in which $|\Delta v|$ was greater than that of the actual data.

\setcounter{figure}{0}
\makeatletter
\renewcommand{\thefigure}{S\@arabic\c@figure}
\onecolumngrid
\section{Supplementary figures}

\begin{figure*}[h]
\begin{center}
\includegraphics[height=3.7in]{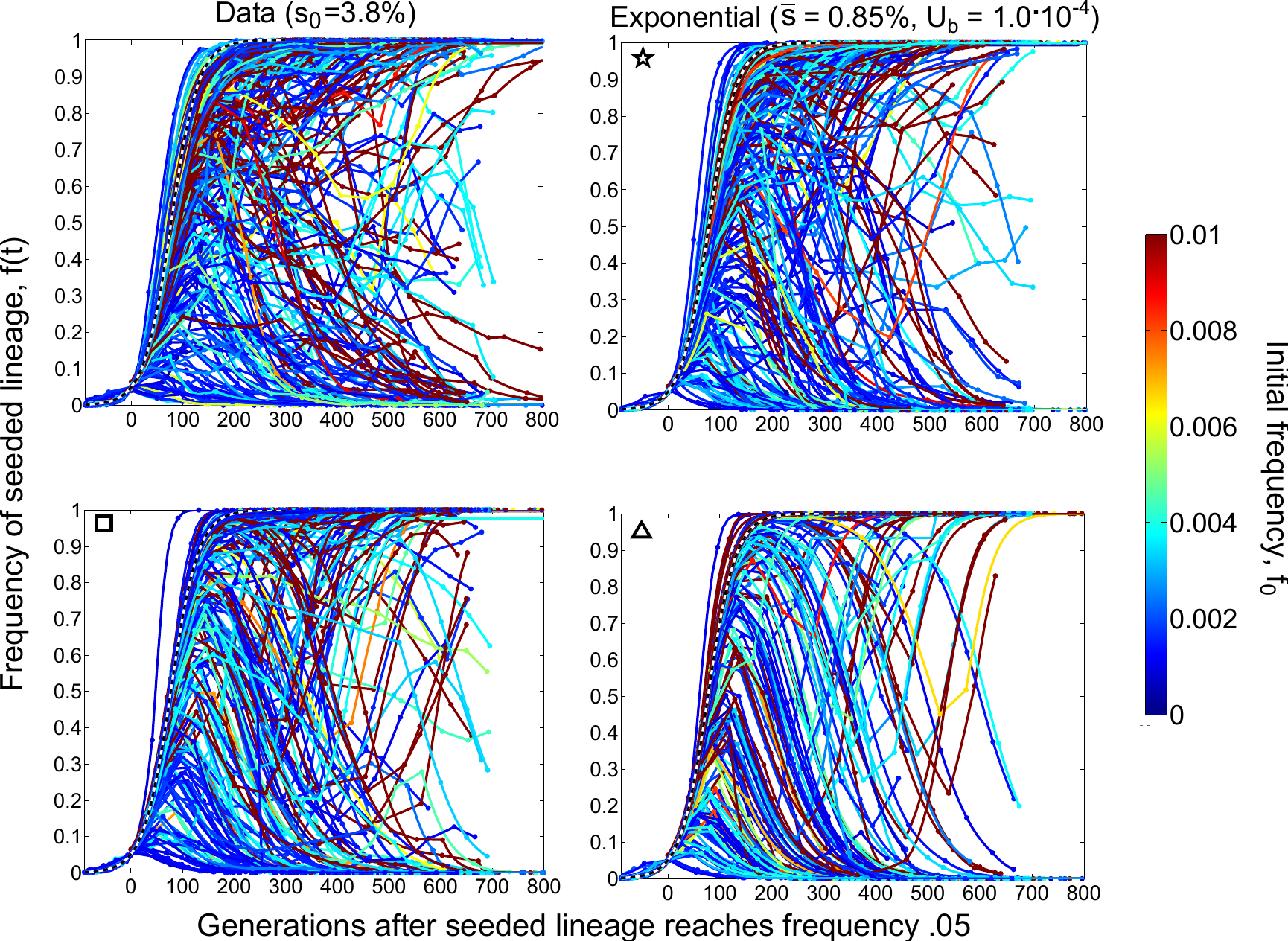}
\caption{{\bf Lineage dynamics data and simulations for $\so = 3.8\%$.} Each panel shows the trajectories of seeded lineages with initial fitness $\so=3.8\%$ as observed in the experiment (top left) and as reproduced by simulations assuming the DFE parameters indicated above each panel. \label{fig:linDynamicsDatAndSims2}}
\centering
\end{center}
\end{figure*}

\begin{figure*}[h]
\begin{center}
\includegraphics[height=3.7in]{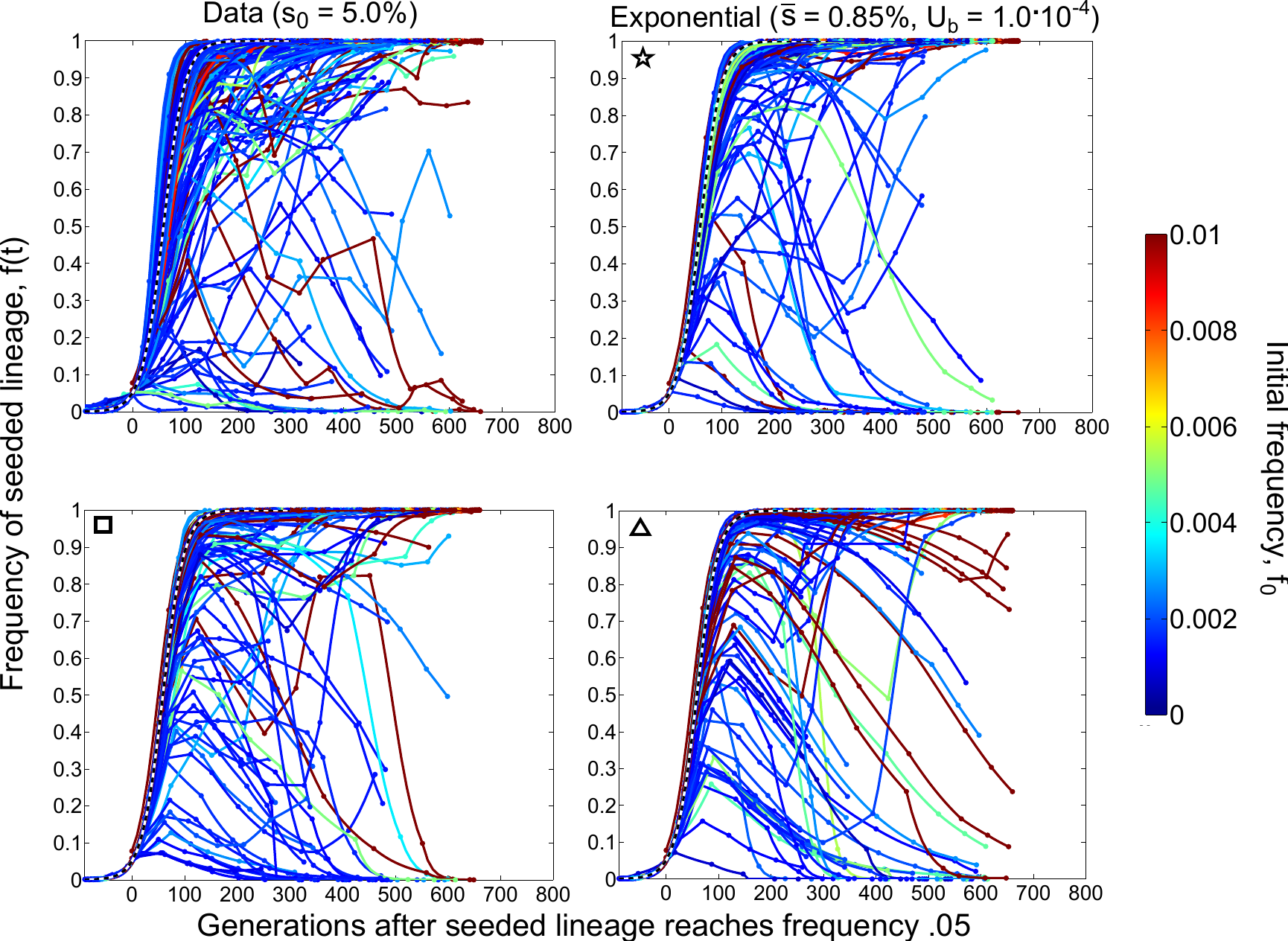}
\caption{{\bf Lineage dynamics data and simulations for $\so = 5.0\%$.} Each panel shows the trajectories of seeded lineages with initial fitness $\so=5.0\%$ as observed in the experiment (top left) and as reproduced by simulations assuming the DFE parameters indicated above each panel. \label{fig:linDynamicsDatAndSims3}}
\centering
\end{center}
\end{figure*}

\begin{figure*}[h]
\begin{center}
\includegraphics[height=3.7in]{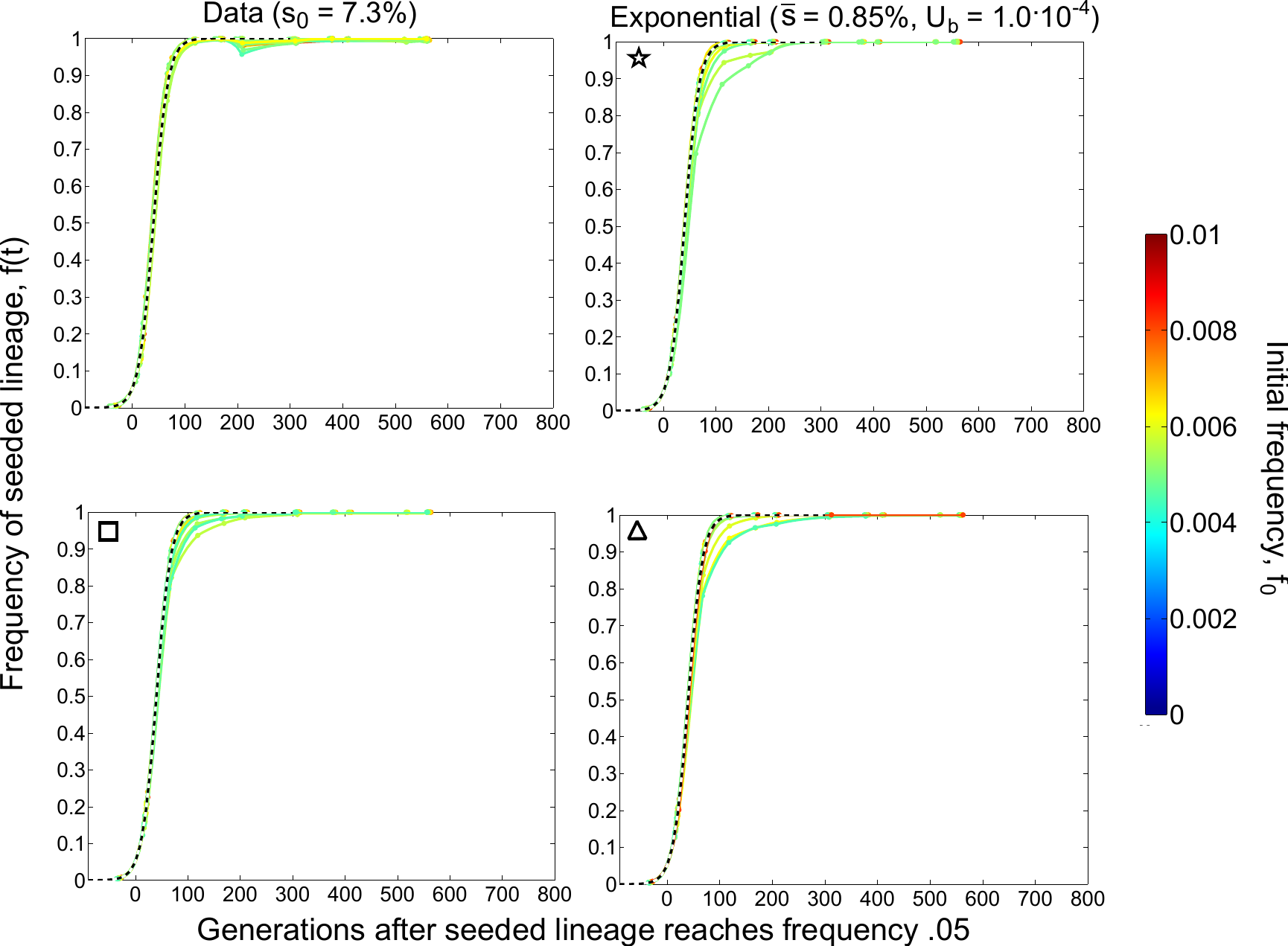}
\caption{{\bf Lineage dynamics data and simulations for $\so = 7.3\%$.} Each panel shows the trajectories of seeded lineages with initial fitness $\so=7.3\%$ as observed in the experiment (top left) and as reproduced by simulations assuming the DFE parameters indicated above each panel (which are also the ones indicated by the star, triangle and square in \fig{fig:DFEsInferred}). \label{fig:linDynamicsDatAndSims4}}
\centering
\end{center}
\end{figure*}

\begin{figure}
\begin{center}
\includegraphics[width=3.4in]{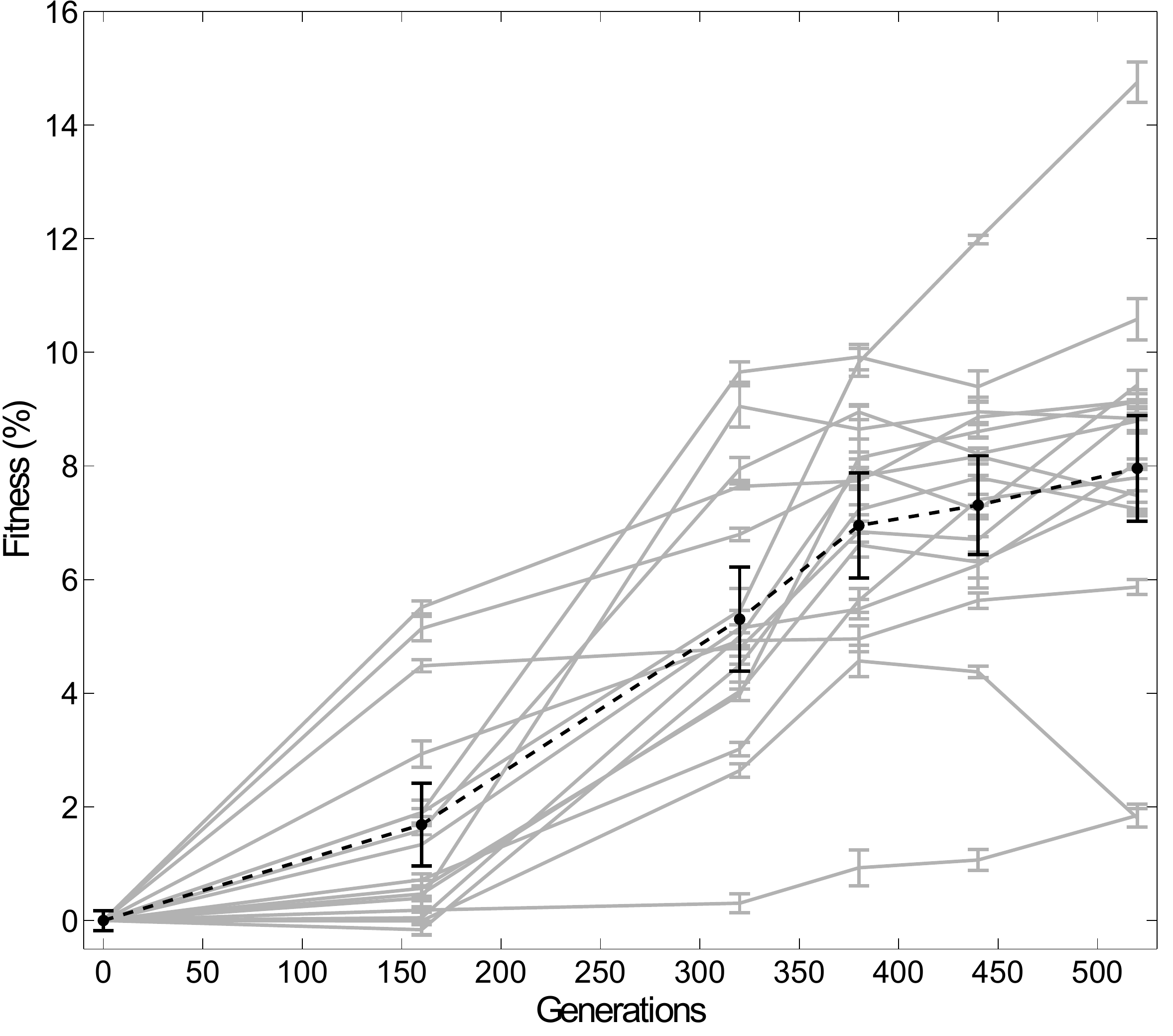}
\caption{{\bf The rate of adaptation.} The fitness over time of 16 experimental control populations (grey curves) and their mean (black curve). The error bars are $\pm$1~s.e.m. \label{fig:adapRateDatVsimSI}}
\centering
\end{center}
\end{figure}

\end{document}